# Thermal conductivity minimum of graded superlattices due to phonon localization

Yangyu Guo [1*], Marc Bescond [2], Zhongwei Zhang [1], Shiyun Xiong [1], Kazuhiko Hirakawa [1], Masahiro Nomura [1†], Sebastian Volz [1,2‡]

[1] *Institute of Industrial Science, The University of Tokyo, Tokyo 153-8505, Japan*
[2] *LIMMS, CNRS-IIS UMI 2820, The University of Tokyo, Tokyo 153-8505, Japan*

(Dated: Apr. 23, 2021)

## Abstract

The Anderson localization of thermal phonons has been shown only in few nanostructures with strong random disorder by the exponential decay of transmission to zero and a thermal conductivity maximum when increasing system length. In this work, we present a path to demonstrate the phonon localization with distinctive features in graded superlattices with short-range order and long-range disorder. A thermal conductivity minimum with system length appears due to the exponential decay of transmission to a non-zero constant, which is a feature of partial phonon localization caused by the moderate disorder. We provide clear evidence of localization through the combined analysis of the participation ratio, transmission, and real-space phonon number density distribution based on our quantum transport simulation. The present work would promote heat conduction engineering by localization via the wave nature of phonons.

[*] yyguo@iis.u-tokyo.ac.jp
[†] nomura@iis.u-tokyo.ac.jp
[‡] volz@iis.u-tokyo.ac.jp

# 1. Introduction

Heat conduction engineering in semiconductor micro- and nanostructures has been widely investigated in the past decades based on the particle picture of phonons [1-5]. As phonons are essentially wave-packets with finite width, the wave picture provides another important degree to engineer heat conduction when the characteristic size of nanostructures becomes comparable to the coherence length of phonons [6-9]. In recent years, there are mainly two categories of nanostructures proposed for the heat conduction tuning via the wave nature of phonons: nano-phononic meta-materials (pillared thin films [10, 11], nanowire cages [12], *etc.*) and nano-phononic crystals (superlattices (SLs) [13-17], periodic porous nanostructures [18-20], *etc.*) based on resonance and band-folding respectively.

Localization is a well-known wave phenomenon that significantly impedes transport since the pioneering work of Anderson [21]. It denotes the absence of wave diffusion in a disordered medium due to the coherent backscattering and destructive interference [22], and has been widely studied with electrons [23], electromagnetic waves [24], and elastic waves [25]. The localization of thermal phonons, induced by the disorder of bond strength and/or atomic mass, actually widely exists in amorphous materials [26] and alloys [27, 28]. However, the localization effect on heat transport is invisible or unknown in these systems.

The phonon transmission and localization in disordered SLs systems were studied based on elastic continuum models in the low-frequency limit [29-31]. The influence of phonon localization on thermal transport properties has been more extensively explored by atomistic simulation methods via either ballistic non-equilibrium Green's function (NEGF) [32-35] or molecular dynamics (MD) [36-40]. The localization effect was shown to be not observable in thermal conductivity of isotope-disordered nanotubes due to the small contribution of high-frequency localized phonons at temperatures low enough to preserve phonon coherence from anharmonic scattering [32, 33]. Important progress was made to explicitly demonstrate the phonon localization by the thermal conductivity maximum of periodic SLs with randomly embedded nanodots [34], which has been experimentally observed recently [41]. The decrease of thermal conductivity with total length is attributed to the exponential decay of phonon transmission to zero [32-34], which is more directly evidenced in a recent modal-level NEGF study of randomly aperiodic SLs [35]. The phonon localization in aperiodic SLs in the presence of anharmonicity and interfacial mixing is thoroughly studied by MD simulation [38], where the thermal conductivity maximum appears at a sufficient level of periodicity disorder. There are also some MD studies of phonon localization in random multilayer systems [36, 40], disordered porous nanophononic crystals [37], and heterogenous interfaces with atomic inter-diffusion [39]. However, the thermal conductivity maximum is only obtained in the coherent part [36, 37], while any measurable evidence of phonon localization remains still ambiguous in these systems.

To sum up, the previous works on phonon localization in heat conduction are mainly featured by: (i) random disorder, (ii) exponential decay of phonon transmission to zero with increasing system length, and (iii) a thermal conductivity maximum with length. The feature (ii) is a signature of strong localization [42] caused by feature (i) and finally leads to feature (iii) [34]. In this work, through a ballistic NEGF simulation, we demonstrate phonon localization in graded SLs with very different features: (i) short-range order and long-range disorder, (ii) exponential decay of phonon transmission with system length to a non-zero constant, and (iii) a thermal conductivity minimum with length. This kind of localization is not strong as the level of disorder is not sufficiently high. As a result, the phonon transmission does not fully decay to zero with increasing system length. The decay of transmission to a non-zero constant was actually shown in few systems recently [37, 38]. In this work, we aim to draw attention to such partial localization [38] of broad-band phonons, which was seldomly considered in heat conduction.

The remainder of the article is organized as follows. The physical model of graded Si/Ge SLs and the corresponding NEGF methodology will be introduced in Section 2. The results and discussions of phonon localization in graded SLs system will be given in Section 3. Finally, the concluding remarks will be made in Section 4.

**2. Physical model and Methodology**

In this section, firstly we explain the detailed configurations of graded Si/Ge SLs in Section 2.1. In Section 2.2, we introduce the ballistic phonon NEGF formalism, together with the calculation of several important variables from the Green's functions.

*2.1 Physical model*

We design a series of graded Si/Ge SLs as exemplified in Figure 1(a), with a reference periodic Si/Ge SLs shown in Figure 1(b). In both the graded and periodic SLs, the interfaces are atomically smooth. The graded SLs are made up of several different SLs units, with each SLs unit containing Np periods of each period length. In the example shown in Figure 1(a), there are 3 SLs units with the period lengths: $p_1 = 1$ uc, $p_2 = 2$ uc, $p_3 = 3$ uc, and the number of periods Np = 2 in each SLs unit. Here 1 uc denotes the length of one conventional unit cell of Si. Other configurations with more SLs units are given in Table 1. The reference periodic SLs in Figure 1(b) have a period length of p = 1 uc. The present graded SLs have both short-range order and long-range disorder. With increasing number of SLs units, the level of disorder elevates. For graded SLs with a specific number of SLs units, the total system length *L* increases as Np increases. For simplicity, we assume the lattice constant and force constant of Ge to be the same as those of Si, with only the atomic mass difference taken into account [43, 44]. All the considered period lengths lie within a range of about 1 ~ 4 nm, which are comparable to the dominant phonon coherence length in Si/Ge SLs [38, 45] and ensure the coherent condition for phonon localization.

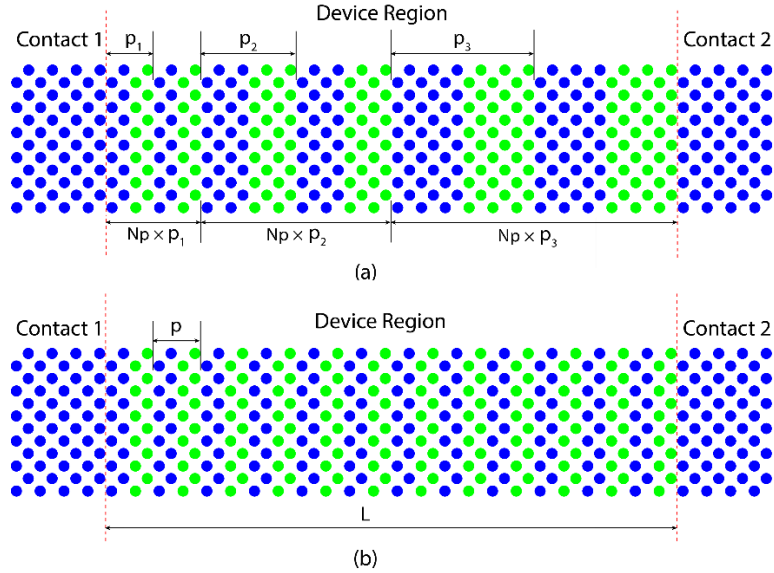

Figure 1. Schematic of the physical model: (a) graded Si/Ge superlattices (SLs) made up of 3 SLs units with 3 period-lengths separately: $p_1 = 1$ uc, $p_2 = 2$ uc, $p_3 = 3$ uc, and Np periods in each SLs unit (Np = 2), other configurations of graded SLs are detailed in Table 1; (b) periodic Si/Ge SLs with a period length of p = 1 uc. The blue and green spheres represent Si and Ge atoms,

respectively. The transverse direction is periodic. Here 1uc denotes the length of one conventional unit cell of Si. The system length is denoted by *L*.

Table 1. Graded Si/Ge superlattices (SLs) with different numbers of SLs units and the corresponding period lengths. Here 1 uc denotes the length of one conventional unit cell of Si.

| Number of SLs units | Period lengths |
| --- | --- |
| 3 | p = 1 uc, 2 uc, 3 uc |
| 4 | p = 1 uc, 2 uc, 3 uc, 4 uc |
| 5 | p = 1 uc, 2 uc, 3 uc, 4 uc, 5 uc |
| 6 | p = 1 uc, 2 uc, 3 uc, 4 uc, 5 uc, 6 uc |
| 7 | p = 1 uc, 2 uc, 3 uc, 4 uc, 5 uc, 6 uc, 7 uc |

*2.2 NEGF methodology*

As phonon localization in heat conduction is coherent in nature, we apply the ballistic NEGF formalism [46-48], with the retarded Green's function of the device region (graded or periodic SLs here) expressed as:

$$\mathbf{G}^{\text{R}}(\omega;\mathbf{q}_\perp) = \left[\omega^2 \mathbf{I} - \tilde{\mathbf{\Phi}}(\mathbf{q}_\perp) - \mathbf{\Sigma}^{\text{R}}(\omega;\mathbf{q}_\perp)\right]^{-1}, \qquad (1)$$

with **I** the unity matrix, and $(\omega;\mathbf{q}_\perp)$ denote the frequency and wave vector dependences along the transport and the periodic transverse direction, respectively, $\tilde{\mathbf{\Phi}}(\mathbf{q}_\perp)$ being the Fourier's representation of the harmonic force constant matrix [44, 48]. The retarded self-energy matrix only includes the contribution from two contacts: $\mathbf{\Sigma}^{\text{R}}(\omega;\mathbf{q}_\perp) = \mathbf{\Sigma}_1^{\text{R}}(\omega;\mathbf{q}_\perp) + \mathbf{\Sigma}_2^{\text{R}}(\omega;\mathbf{q}_\perp)$, which are related to the surface Green's function of contacts computed by the decimation technique [49]. The greater/lesser Green's function is computed by [44, 50]:

$$\mathbf{G}^{>,<}(\omega;\mathbf{q}_\perp) = \mathbf{G}^{\text{R}}(\omega;\mathbf{q}_\perp)\mathbf{\Sigma}^{>,<}(\omega;\mathbf{q}_\perp)\mathbf{G}^{\text{A}}(\omega;\mathbf{q}_\perp), \qquad (2)$$

where the advanced Green's function $\mathbf{G}^{\text{A}}$ is the Hermitian conjugate of $\mathbf{G}^{\text{R}}$, and the greater/lesser self-energy matrix also only includes the contribution from the two contacts: $\mathbf{\Sigma}^{>,<}(\omega;\mathbf{q}_\perp) = \mathbf{\Sigma}_1^{>,<}(\omega;\mathbf{q}_\perp) + \mathbf{\Sigma}_2^{>,<}(\omega;\mathbf{q}_\perp)$, which are related to the retarded contact self-energy matrices and Bose-Einstein equilibrium distributions [44]. Note that in ballistic NEGF, it is usually not necessary to calculate $\mathbf{G}^{>,<}$, which are required instead in the present work to compute the local phonon number density as introduced later.

Once the Green's functions are resolved, the transmission through the SLs is obtained based on the Caroli formula [46-48]:

$$\Xi(\omega) = \frac{1}{N}\sum_{\mathbf{q}_\perp}\mathrm{Tr}\left[\mathbf{\Gamma}_1(\omega;\mathbf{q}_\perp)\mathbf{G}^R(\omega;\mathbf{q}_\perp)\mathbf{\Gamma}_2(\omega;\mathbf{q}_\perp)\mathbf{G}^A(\omega;\mathbf{q}_\perp)\right], \tag{3}$$

where $N$ is the number of transverse wave vectors, 'Tr' denotes the trace of a square matrix, and the broadening matrix is defined as: $\mathbf{\Gamma}_{1(2)} = i\left(\mathbf{\Sigma}^R_{1(2)} - \mathbf{\Sigma}^A_{1(2)}\right)$, with the advanced contact self-energy matrix $\mathbf{\Sigma}^A$ being the Hermitian conjugate of $\mathbf{\Sigma}^R$. The thermal conductance of the SLs is calculated based on the Landauer's formula as [46-48]:

$$\sigma = \frac{1}{s}\int_0^\infty \frac{\hbar\omega}{2\pi}\Xi(\omega)\frac{\partial f_{\mathrm{BE}}(\omega)}{\partial T}d\omega, \tag{4}$$

where $s$ is the area of transverse cross-section, and $f_{\mathrm{BE}}$ the Bose-Einstein distribution. The thermal conductivity of the SLs is related to its thermal conductance as: $\kappa = \sigma L$. We also introduce the normalized thermal conductivity accumulation function versus frequency:

$$\frac{\kappa_{\mathrm{accum}}(\omega)}{\kappa} = \frac{1}{\sigma s}\int_0^\omega \frac{\hbar\omega'}{2\pi}\Xi(\omega')\frac{\partial f_{\mathrm{BE}}(\omega')}{\partial T}d\omega', \tag{5}$$

which provides a direct visualization of the frequency range where phonon localization takes place.

The local phonon number density is related to the diagonal blocks of $\mathbf{G}^<$ as [44, 51]:

$$\rho(\varepsilon,\mathbf{R}_n) = \mathrm{Tr}\left[\frac{1}{N}\sum_{\mathbf{q}_\perp}i\mathbf{G}^<_{nn}(\omega;\mathbf{q}_\perp)\right], \tag{6}$$

where $\mathbf{R}_n$ is the position of atomic site $n$, and $\varepsilon \equiv \omega^2$ is the eigen-value of harmonic force constant matrix, with the relation: $\rho(\omega,\mathbf{R}_n) = \rho(\varepsilon,\mathbf{R}_n)\omega/\pi$. The local phonon number density provides an intuitive real-space representation of phonon localization: the phonon number density of localized states is concentrated around some atomic sites, as to be shown later. The local density of states (LDOS) is defined as [44, 51]:

$$\mathrm{LDOS}(\varepsilon,\mathbf{R}_n) = \mathrm{Tr}\left[\mathbf{A}_{nn}(\omega)\right], \tag{7}$$

with the spectral function matrix computed by:

$$\mathbf{A}_{nn}(\omega) = \frac{1}{N}\sum_{\mathbf{q}_\perp}\mathbf{A}_{nn}(\omega;\mathbf{q}_\perp) = \frac{1}{N}\sum_{\mathbf{q}_\perp}i\left[\mathbf{G}^>_{nn}(\omega;\mathbf{q}_\perp) - \mathbf{G}^<_{nn}(\omega;\mathbf{q}_\perp)\right]. \tag{8}$$

Also, we have the relation for LDOS as: $\mathrm{LDOS}(\omega,\mathbf{R}_n) = \mathrm{LDOS}(\varepsilon,\mathbf{R}_n)\omega/\pi$. The overall DOS of the system is calculated as:

$$\text{DOS}(\omega) = \frac{1}{N_a} \sum_n \text{LDOS}(\omega, \mathbf{R}_n), \tag{9}$$

with $N_a$ the total number of atoms in the system. Once we obtain the LDOS throughout the system, the spectral participation ratio (PR) is introduced to characterize the level of phonon localization [52, 53]:

$$\text{PR}(\omega) = \frac{\left[\sum_n \text{LDOS}(\omega, \mathbf{R}_n)\right]^2}{N_a \sum_n \left[\text{LDOS}(\omega, \mathbf{R}_n)\right]^2}. \tag{10}$$

To account for the influence of the number of states at a specific frequency, we also introduce a weighted PR defined as the product of PR and DOS [53].

In terms of the numerical implementation, we adopt a recursive algorithm to solve the retarded and greater/lesser Green's functions in Eq. (1) and Eq. (2) within a massively parallelized framework [44]. The harmonic force constants of Si are obtained from the first-principle calculation, with the details given in our previous work [44]. The optimized size of one unit cell of Si is 1 uc = 5.4018 Å. A mesh of $N_m$ = 200 for the frequency points and of $N$ = 20 × 20 for the transverse wave vector points are adopted after careful independence check.

## 3. Results and Discussions

In this section, firstly we demonstrate the appearance of a thermal conductivity minimum with the length of the graded Si/Ge SLs due to phonon localization in Section 3.1. A detailed analysis of the length-dependent transmission and participation ratio is presented as the evidence of phonon localization in Section 3.2. Finally, the real-space localization pattern is shown via the phonon number density distribution.

*3.1 Thermal conductivity minimum*

Firstly, the results of thermal conductance and conductivity of periodic Si/Ge SLs with p = 1 uc at 200 K are given in Figure 2 as a reference for later discussion. The reason of considering the temperature of 200K will be given later. With increasing number of periods, the thermal conductance decreases and rapidly converges to a constant value after about 4 ~ 5 periods as shown in Figure 2(a). This is due to the band-folding process in SLs, where several periods are required for the reflections and interferences of lattice waves to form the SLs eigen-modes [43]. Once the folded modes are well formed, they transport ballistically across the SLs as in a homogenous medium with a conductance independent of the system length [14]. The thermal conductivity of periodic SLs increases almost linearly with the system length as shown in Figure 2(b), which also indicates ballistic heat transport. The present trend of thermal conductance and conductivity for the periodic SLs with smooth interfaces agrees quite well with that in previous NEGF studies [43, 54].

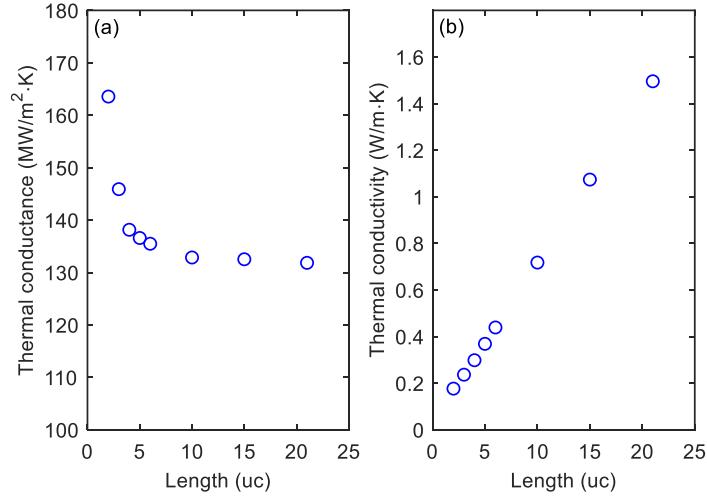

Figure 2. Length-dependent thermal conductance (a) and conductivity (b) of periodic Si/Ge superlattices with a period p = 1 uc at 200 K.

Then we discuss the results of thermal conductance and conductivity of graded Si/Ge SLs at 200 K in Figure 3. For all the graded SLs with different numbers of SLs units, generally the thermal conductance rapidly decreases with total length and gradually converges at Np ≥ 4 ~ 5, as shown in Figure 3(a). The decrease of thermal conductance is more significant compared to the periodic SLs case. With increasing number of SLs units, the thermal conductance converges to lower values. This is caused by the elevating level of long-range disorder in the graded SLs, which induces elastic backscattering and destructive interference of phonon wave-packets. As the number of SLs units increases from 3 to 5, the elastic backscattering becomes stronger to approach the quantum diffusion limit ($\sigma \sim 1/L$) [21, 22] at the initial stage (Np ≤ 4 ~ 5), as indicated in the double-log-scale inset in Figure 3(a). As a result, the thermal conductivity of the graded SLs transits from an increasing trend with length (the case of 3 and 4 SLs units) to a length-independent regime (at Np ⩽ 4 ~ 5 in the case of 5 SLs units), as it is demonstrated in Figure 3(b). As the number of SLs units further increases, the elastic backscattering is so strong that the decrease of thermal conductance with length at the initial stage is faster than the quantum diffusion limit, *i.e.* $\sigma \sim 1 / L^{\alpha}$ with $\alpha > 1$. Thus, the thermal conductivity even decreases with length (at Np ≤ 4 ~ 5 in the case of 6 and 7 SLs units) as shown in Figure 3(b), which is a signature of phonon localization. For all the cases, the thermal conductivity starts to increase with length at Np ≥ 4 ~ 5 since the thermal conductance gradually converges. Therefore, a thermal conductivity minimum with length appears at Np = 4 ~ 5 for the graded SLs with 6 and 7 SLs units. The minimum shall come from a competition between the phonon localization and ballistic transport. As the level of disorder in the graded SLs is moderate compared to the random disorder (*e.g.* in aperiodic SLs [35, 38]), a strong localization [42] cannot be reached and the ballistic transport wins out at Np ≥ 4 ~ 5. As a comparison, the thermal conductivity maximum reported in the strong localization regime [34, 35, 38, 41] results from the ballistic-to-localized transition as the system length increases to be comparable to and larger than the phonon localization length. The present minimum is also different from the previous thermal conductivity minimum [13, 15-17] due to the coherent-to-incoherent transition when increasing the period length of SLs under a fixed total length. The heat transport in graded SLs is somehow in a *partial localization regime*, which was introduced for the high-frequency phonons in aperiodic SLs due to decoherence as the period length is comparable to or even larger than the coherence length of those phonons [38]. We will provide deeper analysis and more direct evidence of localization via the phonon PR, transmission, and number density distribution in the sub-sections as follows.

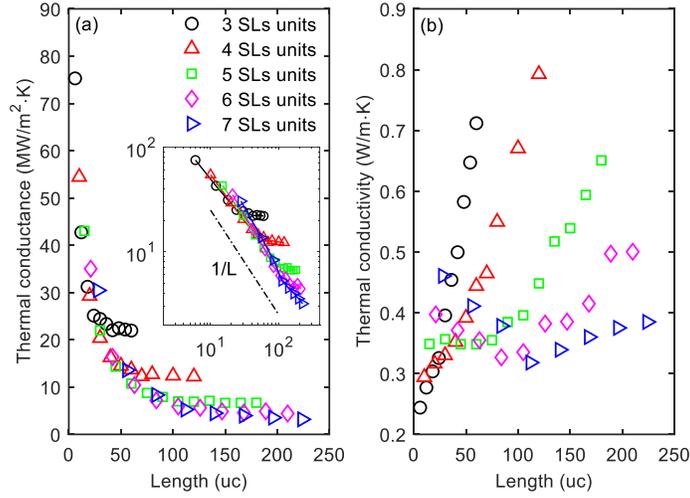

Figure 3. Length-dependent thermal conductance (a) and conductivity (b) of graded Si/Ge superlattices (SLs) with different numbers of SLs units at 200 K. The inset in (a) shows the results in double-log scale, where the dash-dotted line indicates the scale of quantum diffusion: $\sigma \sim 1/L$, with $L$ denoting the system length. A thermal conductivity minimum with length starts to appear in (b) as the level of disorder (number of SLs units) increases. In both (a) and (b), the length increases as Np increases from 1. The unit of length is 1 uc = 5.4018 Å.

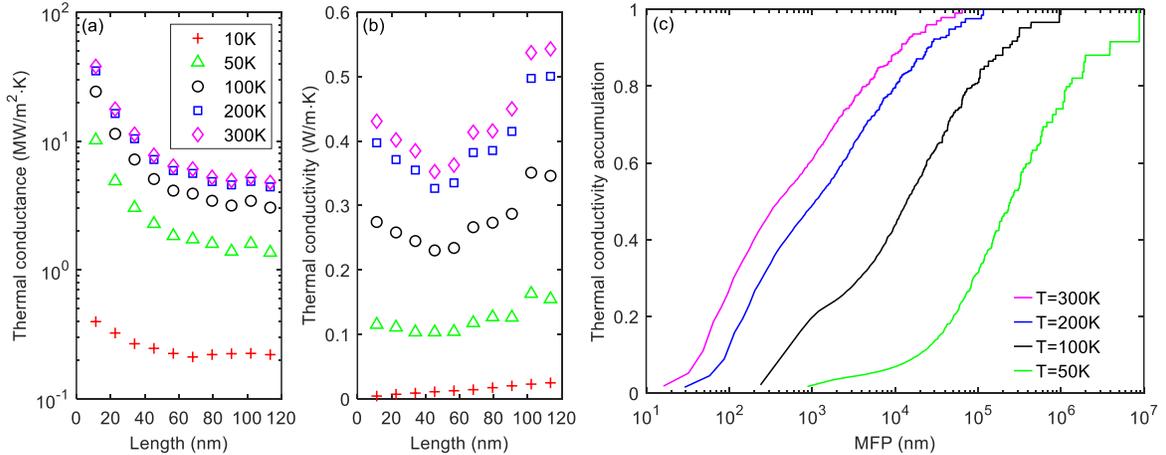

Figure 4. Temperature dependence of the length-dependent thermal conductance (a) and conductivity (b) of graded Si/Ge superlattices (SLs) with 6 SLs units, where the length increases as Np increases from 1; (c) shows the normalized thermal conductivity accumulation versus mean free path (MFP) of bulk Si at different temperatures, calculated by first-principle method with a phonon **q**-mesh of $64 \times 64 \times 64$.

The temperature dependence of the thermal conductance and conductivity of the graded Si/Ge SLs with 6 SLs units is plotted in Figure 4(a) and Figure 4(b) respectively. As shown in Figure 4(a), the thermal conductance decreases with decreasing temperature. The dip of thermal conductivity minimum with length becomes smaller as the temperature is lower, and fully vanishes at 10 K, as shown in Figure 4(b). This indicates that the phonon localization is mainly contributed by modes in the moderate- and/or high-frequency range, as to be uncovered later. To ensure the population of these localized phonon modes, the system temperature cannot be too low. On the other hand, the system temperature shall be sufficiently low to suppress the anharmonic phonon-phonon scattering which will destroy the phonon coherence. To sum up, to observe phonon localization, the system temperature can be neither too low nor too high, which is similar to the condition in the periodic SLs with embedded nanodots [34, 41]. To estimate the effect of anharmonic scattering missing in our ballistic NEGF simulation, we calculate the phonon mean free path (MFP) distribution of bulk Si at different temperatures by the first-principle method [55]. As shown by the normalized thermal conductivity accumulation function versus MFP in Figure 4(c), the phonons with MFP < 100 nm contribute to only ~ 10% and nearly 0% of the total thermal conductivity at 200 K and 100 K, respectively. Considering that the thermal conductivity minimum takes place at a system length < 100 nm, the temperature range of 100 ~ 200 K with negligible anharmonic scattering shall be good for the observation of phonon localization. Throughout the present work, we mainly consider and discuss the simulation results at 200 K.

Finally, to quantify the effect of phonon localization, we provide a comparison of the thermal conductance of graded SLs to that of the reference periodic SLs with p = 1 uc in Table 2. The length of the graded SLs is set to Np = 4 where the thermal conductivity minimum takes place. The length of the periodic SLs is the same as that of graded SLs with each specific number of SLs units. The thermal conductance of the periodic SLs is shown to be almost independent of the total length in the present range, which is consistent with the trend in Figure 2(a). In contrast, the thermal conductance of graded SLs is reduced a lot due to phonon localization. The reduction ratio increases with the number of SLs units, and reaches as high as ~ 96% at 7 SLs units. Therefore, the graded SLs is a good system to achieve low thermal conductivity. Although its performance might be slightly lower than that of the random multilayer structures [36] or optimized aperiodic SLs [53], the graded SLs seem to be more easily fabricated experimentally due to the short-range order.

Table 2. Thermal conductance of the graded superlattices (SLs) with different numbers of SLs units (with Np = 4 in each SLs unit), and the reference periodic SLs with p = 1 uc and the same corresponding length at 200 K. The unit of length is 1 uc = 5.4018 Å.

| Number of SLs units | 3 | 4 | 5 | 6 | 7 |
|---|---|---|---|---|---|
| Length (uc) | 24 | 40 | 60 | 84 | 112 |
| Periodic SLs (MW/m$^2$·K) | 131.74 | 131.05 | 132.42 | 130.56 | 132.17 |
| Graded SLs (MW/m$^2$·K) | 25.11 | 16.30 | 10.75 | 7.20 | 5.26 |
| Reduction (%) | 80.94 | 87.56 | 91.88 | 94.49 | 96.02 |

*3.2 Length-dependent participation ratio and transmission*

To illustrate the phonon localization, we show the length-dependent PR, weighted PR and transmission of the graded SLs with 6 SLs units in Figure 5. The PR, weighted PR and the transmission all decrease with increasing length and converge after L ~ 84 uc (Np ~ 4). In comparison to the reference periodic SLs with p = 1 uc, the PR of the moderate-frequency (3 ~ 7 THz) and high-frequency (> ~ 10 THz) phonons in graded SLs is reduced considerably, which is an evidence of localization. The reduction of PR of very low-frequency phonons (< ~ 1 THz) is relatively smaller, indicating ballistic heat transport. The convergence length Np ~ 4 is well consistent with the length where the thermal conductance converges and the thermal conductivity minimum takes place in Section 3.1. Since there are 6 different SLs units in the graded SLs, the band-folding in each SLs unit may induce flat bands (localized modes) and band gaps. To distinguish this kind of localization and that induced by the destructive interference of folded eigen-modes among different SLs units, we compare the PR, weighted PR and transmission of 6 periodic SLs with each period length (p = 1 ~ 6 uc) in the graded SLs to those of the graded SLs in Figure 6. The length of the periodic SLs is fixed to 60 uc, which contains a sufficient number of periods to ensure well-formed band-folding in all the cases. With increasing period length in the periodic SLs, the high-frequency phonons are more strongly localized due to flat bands and become almost fully localized at p = 5 uc and p = 6 uc as shown in Figure 6(a) and (b), which shall be the dominant cause of localization in the high-frequency range in the graded SLs. However, for the moderate-frequency phonons, the PR and weighted PR in the periodic SLs are more or less independent of the period length. Once the different SLs units are connected to constitute the graded SLs, the PR and weighted PR are much reduced and smaller than any of those of the periodic SLs. The reduction shall come from the interference of folded modes in different SLs units, which introduces extremely strong extinction of transmission in the moderate-frequency range as it is shown in Figure 6(c). The convergence length of Np ~ 4 represents the number of periods that is required in each SLs unit to well form the corresponding folded eigen-modes. Once the folded modes in each SLs unit are well formed and interfere destructively, the extent of localization will converge at Np > 4 ~ 5. As the contribution of high-frequency phonons to heat transport is very small in both periodic and graded SLs, we mainly focus on the localization of moderate-frequency phonons in this work.

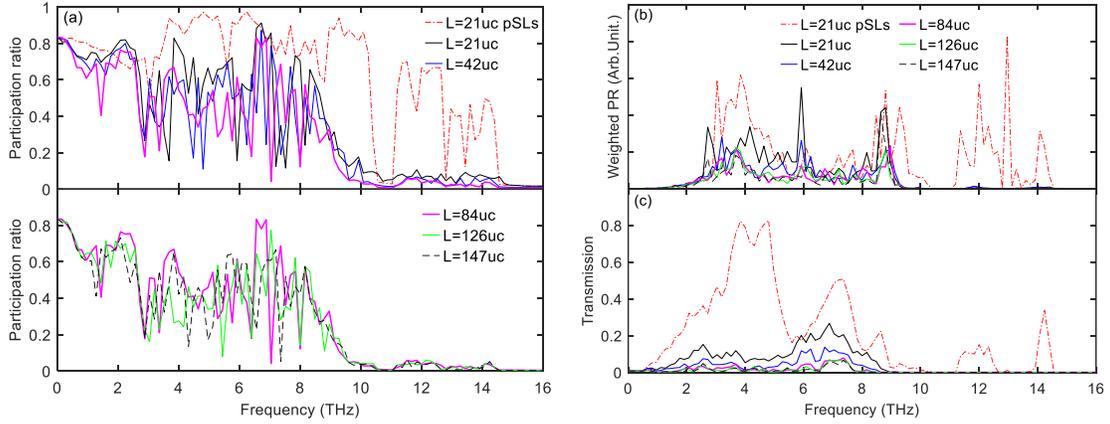

Figure 5. Length-dependent participation ratio (PR) (a), weighted PR (b), and transmission (c) of the graded Si/Ge superlattices (SLs) with 6 SLs units. Figure (a) is divided into two parts to make the plot clear. The length $L$ increases as Np in each SLs unit increases. The weighted PR is defined as the product of PR and DOS (density of states). The dash-dotted red line represents the result of periodic SLs (pSLs) with p = 1 uc as a reference.

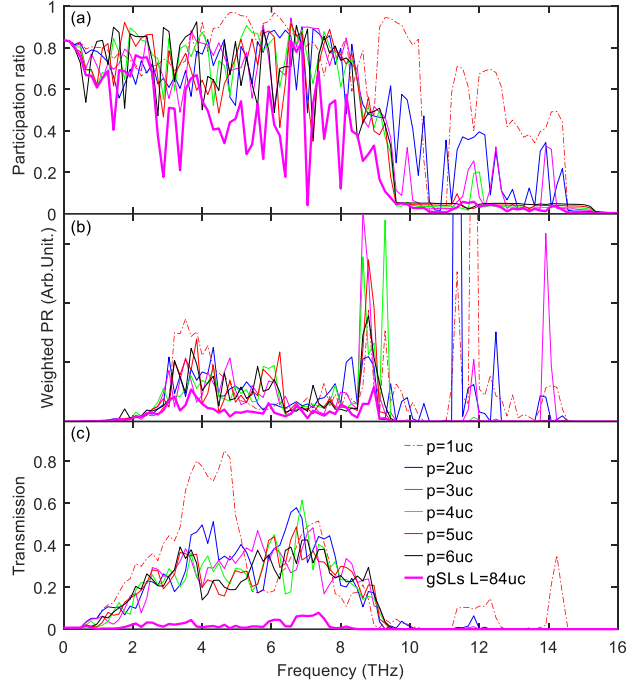

Figure 6. Participation ratio (PR) (a), weighted PR (b), and transmission (c) of periodic Si/Ge superlattices (SLs) with different period lengths and a fixed total length (L = 60 uc). The weighted PR is defined as the product of PR and DOS (density of states). The bold-solid magenta line represents the result of graded SLs (gSLs) with 6 SLs units as a comparison.

The trends of both PR and transmission versus length in the graded SLs with other configurations in Table 1 are rather similar. The PR, weighted PR and transmission at the convergence length (Np = 4) for different configurations are shown in Figure 7(a), (b) and (c) respectively. With increasing number of SL units and level of disorder, the PR, weighted PR and transmission all decrease in the entire spectrum, especially in the moderate-frequency range where the phonon localization is the strongest. The normalized thermal conductivity accumulation function versus frequency in Figure 7(d) provides a corroboration to phonon localization from another point of view. With increasing level of disorder, the slope in the range of 4 ~ 6 THz (the main portion of moderate-frequency range) gradually declines, which represents decreasing contribution to heat transport from the partially localized states. The localization becomes strong enough to make the thermal conductivity decrease with length at 6 and 7 SLs units, where two small plateaux also appear around 1 THz and 6.5 THz.

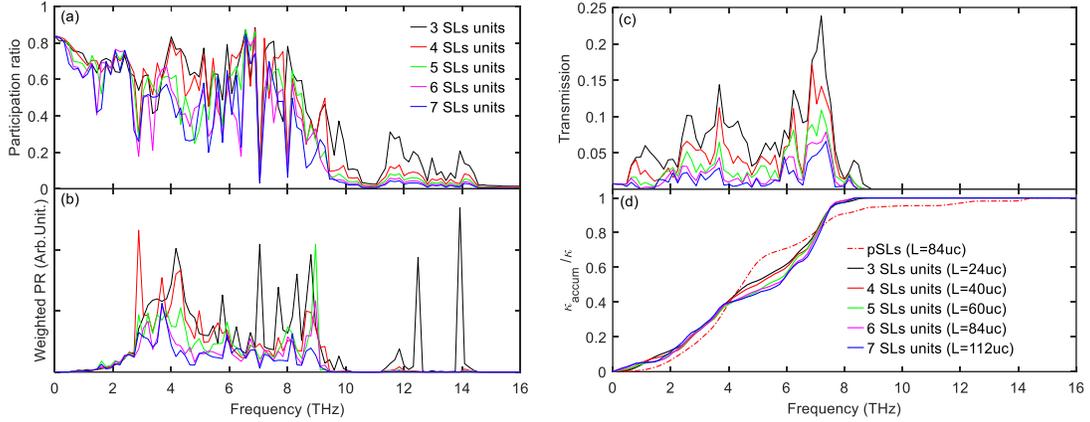

Figure 7. Participation ratio (PR) (a), weighted PR (b), transmission (c), and normalized thermal conductivity accumulation versus frequency (d) of graded Si/Ge superlattices (SLs) with different numbers of SLs units and Np = 4 in each SLs unit. The weighted PR is defined as the product of PR and DOS (density of states). The dash-dotted red line in (d) represents the result of periodic SLs (pSLs) with p = 1 uc as a reference.

Finally, we provide the length-dependent transmission of graded SLs with 6 SLs units for several frequency points in Figure 8. The frequency points are chosen from those with low PR in Figure 5(a). A clear exponential decay of the transmission with length is obtained, which has been widely adopted as an evidence of phonon localization [32-35, 37-39]. However, one salient difference is that the transmission decays to a non-zero constant instead of zero in the strong localization regime [34, 35, 38]. This is a feature of partial localization, which could be caused by either the decoherence of some phonons as highlighted in previous works [37, 38] or the insufficient strength of disorder in the present work. Attributed to the moderate disorder, some portion of phonons at specific frequency could still pass through the graded SLs even if the system length increases continuously. The non-zero constant phonon transmission makes the thermal conductivity of graded SLs increase with length at Np ≥ 4 ~ 5 in Section 3.1. In a recent contribution [38], a correction term considering the phonon decoherence has been added to the conventional exponential decay formula of transmission, which fits well their direct MD simulation results. We do not attempt to fit the length-dependent phonon transmission in Figure 8 by the phenomenological formula [38] as the partial localization is caused by a different mechanism here. A relevant theoretical model is desired to fit the present results in the full range of length and will be the focus of our future work. Comparing to the electron localization that is easier to achieve [42], phonon localization in heat conduction is usually more difficult to observe due to its broad-band feature. Thus, further theoretical study is required on the partial phonon localization, especially its dependence on the strength of disorder.

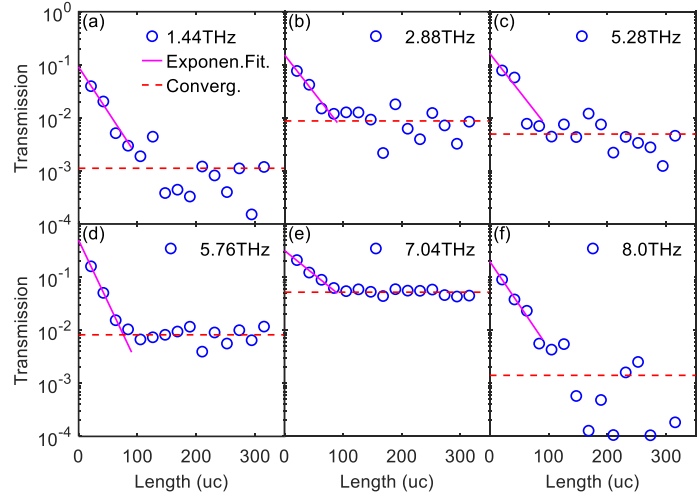

Figure 8. Length-dependent transmission of graded Si/Ge superlattices (SLs) with 6 SLs units at different frequencies. Representative frequency points with low participation ratio (c.f. Figure 5) are shown. The solid magenta line is an exponential fit of the first four length points ($N_p \leq 4$), whereas the dashed red line represents the converged result by averaging the later length points ($N_p \geq 5$). The transmission clearly shows an exponential decay to a non-zero constant with increasing total length.

*3.3 Real-space phonon localization pattern*

To provide a more intuitive picture of phonon localization, we display the phonon number density distribution computed by Eq. (6) for different frequencies in the graded Si/Ge SLs with 6 SLs units and $N_p = 4$ (L = 84 uc) at 200K in Figure 9. The phonon number density distribution has a similar profile to that of the local DOS distribution shown elsewhere [53]; however it also includes the effect of phonon occupation number dependent on the system temperature. For consistent discussion, the considered phonon frequency points are the same as those in Figure 8. For all the considered frequencies except 5.28 THz, the phonon number density is mainly concentrated around a finite spatial region corresponding to a SLs unit, as shown by the shaded area in Figure 9. Two concentrated regions appear at 5.28 THz as shown in Figure 9(c). On the other hand, there are also some small peaks outside the concentrated regions in the number density profile. Those are clear features of localized states (7.04 THz in Figure 9(e)) or partially localized states (other frequencies). In addition, the size of the localization region is roughly inversely proportional to the phonon frequency. The lower-frequency (1.44 THz, 2.88 THz and 5.28 THz) phonons are mainly localized in the SLs units with longer period lengths (p = 4 uc, 3 uc and 5 uc respectively), while the higher-frequency (5.76 THz, 7.04 THz and 8.0 THz) phonons are mainly localized in the SLs units with shorter period lengths (p = 1 uc and 2 uc). This could be explained by the fact the coherence length (width) of phonon wave-packets generally decreases with frequency [45]. In other words, the SLs units with different period lengths in the graded SLs will localize phonons at different frequencies, which introduces significant localization in the broad-band spectrum, especially around the moderate-frequency range.

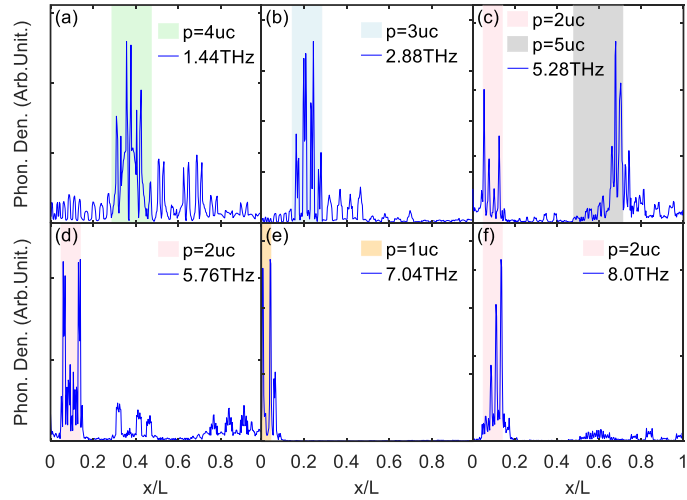

Figure 9. Phonon number density distribution in graded Si/Ge superlattices (SLs) with 6 SLs units and L = 84 uc (Np = 4) at 200 K for different frequencies (c.f. Figure 8). The range of $y$-axis is 0 ~ $1.2\rho_{max}$, with $\rho_{max}$ the maximum local phonon number density inside the graded SLs at the corresponding frequency. The labelled shaded regions represent the SLs units with specific period lengths.

As a comparison, we display the phonon number density distribution for the same set of frequencies in periodic SLs with p = 1 uc and the same total length as that of the graded SLs in Figure 10. The phonon number density is almost uniformly distributed throughout the SLs with fluctuations, which is a clear feature of extended states. As an intermediate case, the graded SLs with 3 SLs units and Np = 4 (L = 24 uc) has a phonon number density profile combining the features of extended states and localized states, as shown in Figure 11. With increasing number of SLs units, the localized states will become more dominant in the graded SLs. However, fully localized states are difficult to reach due to the short-range order in such kind of system, where the partial localization is the main pattern.

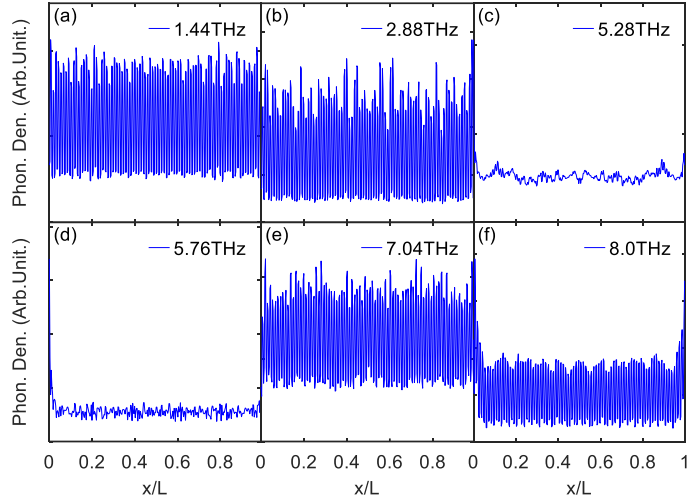

Figure 10. Phonon number density distribution in periodic Si/Ge superlattices with p = 1 uc and L = 84 uc at 200 K for different frequencies. The range of *y*-axis is 0 ~ $1.2\rho_{max}$, with $\rho_{max}$ the maximum local phonon number density inside the superlattices at the corresponding frequency.

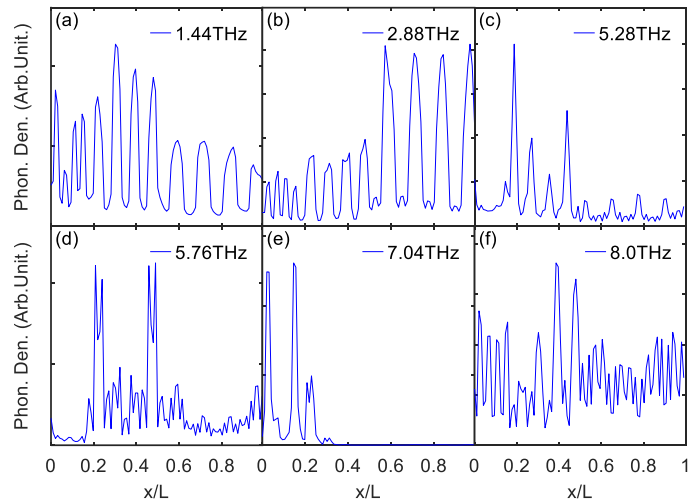

Figure 11. Phonon number density distribution in graded Si/Ge superlattices (SLs) with 3 SLs units and L = 24 uc (Np = 4) at 200 K for different frequencies. The range of *y*-axis is 0 ~ $1.2\rho_{max}$, with $\rho_{max}$ the maximum local phonon number density inside the graded SLs at the corresponding frequency.

Finally, to dig into the localization pattern at the atomic sites, we demonstrate in Figure 12 the contour of the normalized phonon number density distribution within each concentrated region from Figure 9. Within each localization region, there are very few bright zones (for instance, at positions of 30 uc and 32 uc for $\omega = 1.44$ THz in Figure 12(a)) with high phonon number density, which correspond to the sharp peaks of the profiles in Figure 9. These highly localized vibrational states are almost inside the heavier Ge atomic layers (except two cases for $\omega = 5.28$ THz within the considered frequencies around the positions of 4.5 uc and 57 uc in Figure 12(c)). Furthermore, they are located quite close to the Si/Ge interface, which may be caused by the destructive interference of incoming wave-packets and reflected ones from the interface. The present real-space distribution of phonon number density gives a direct picture of phonon localization in the graded Si/Ge SLs.

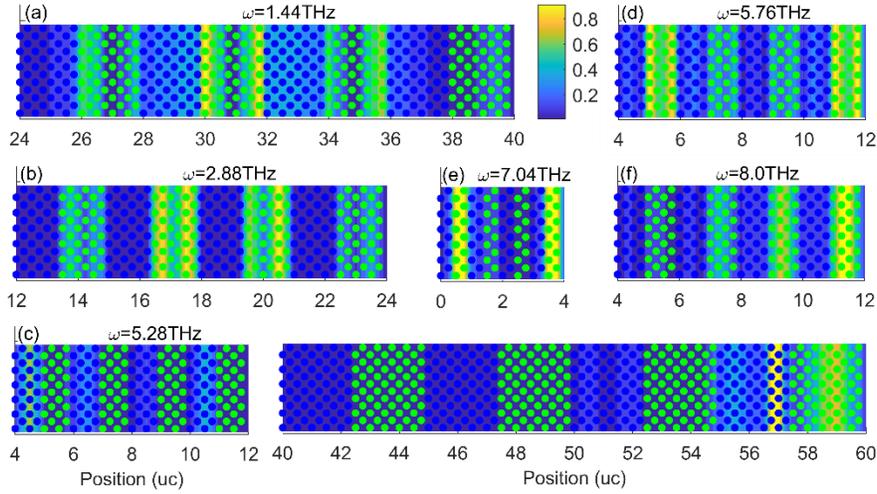

Figure 12. Normalized phonon number density distribution within the corresponding localization superlattices (SLs) unit (c.f. the shaded regions in Figure 9) in the graded Si/Ge SLs with 6 SLs units and L = 84 uc (Np = 4) at 200 K for different frequencies.

## 4. Conclusions

In summary, a novel scheme is presented to demonstrate the phonon localization in designed graded Si/Ge superlattices. With a sufficient level of long-range disorder, a thermal conductivity minimum with system length appears due to the partial localization of moderate-frequency phonons. We illustrate clear evidence of localization via the length-dependent participation ratio and transmission, as well as via the thermal conductivity accumulation function versus frequency. In the partial localization regime, the phonon transmission decays exponentially with length to a non-zero constant. We also provide an intuitive picture of localized states by mapping the concentrated distribution of phonon number density. This work will contribute to a new perspective of phonon localization and also a new avenue to engineering heat conduction via the wave nature of phonons.

**Acknowledgements**

This work was supported by the Postdoctoral Fellowship of Japan Society for the Promotion of Science (P19353), and CREST Japan Science and Technology Agency (JPMJCR19I1 and JPMJCR19Q3), and Grant-in-Aid for Scientific Research S (20H05649). This research used the computational resource of the Oakforest-PACS supercomputer system, The University of Tokyo.

**Data Availability**

The data that support the findings of this study are available from the corresponding author upon reasonable request.